\documentclass[11pt,twoside]{article}
\usepackage[activeacute,spanish]{babel}
\usepackage{baaa-cas}

\usepackage{graphicx}

\usepackage[T1]{fontenc} 

\usepackage{latexsym}
\usepackage{verbatim}

\begin{document}
\def\tablename{Tabla}%

\vskip 1.0cm
\markboth{S. Lipari et al.}%
{Exploding QSOs - Workshop AAA - mayo 2007}

\pagestyle{myheadings}

\vspace*{0.5cm}

\parindent 0pt{ COMUNICACI'ON DE TRABAJO -- CONTRIBUTED  PAPER }

\vskip 0.3cm
\title{The Role of Exploding QSOs in Explosive Models \\
of Evolution, Formation and End of Galaxies}

\author{S. Lipari$^1$, M. Bergmann$^2$, S. F. Sanchez$^3$,
R. Terlevich$^4$, E. Mediavilla$^5$, B.Punsly, B.Garcia$^5$,
W.Zheng$^6$, Y.Taniguchi$^7$, R.Sistero}

\affil{$^1$Observatorio Astronomico de Cordoba and CONICET;
$^2$Gemini Observatory, Chile;
$^3$Calar Alto Observatory, Spain;
$4$Univ. of Cambridge, UK;
$^5$Inst. de Astrof\'{\i}sica de Canarias, Spain;
$^6$Johns Hopkins Univ., USA;
$^7$Tohoku Univ., Japan
}

\begin{resumen}
In this work we analise the role and evidence of exploding
BAL + IR + Fe II QSOs, and
their relation with  new --and previous-- explosive models for
evolution, formation and end of galaxies.
\end{resumen}


\section{ Evolution of Galaxies/QSOs and the Role of Extreme Out Flow/BAL}

Main current issues in astrophysics (from the
theoretical and observational point of view) are the study of
{\it the evolution of IR mergers, IR/BAL QSOs, extreme star formation
processes and the relations between them}.
These issues play an important role
in practically every scenario of formation and evolution of
galaxies and active galactic nuclei (AGNs) (see for references
Lipari \& Terlevich 2006; Sanders \& Mirabel 1996).

On the other hand,
there is increase evidence that extreme galactic out flow (OF) and BAL systems
play a main role at very high redshift (Z $>$ 5.0), i.e. the young universe
(Frye, Broadhurst, Benitez 2002; Maiolino et al. 2003, 2004a,c; Lipari 1994;
Lipari et al. 2005, 2006a,b, 2007; Lipari \& Terlevich 2006).
At low redshift, there are strong evidence of extreme OF,
mainly associated with extreme starburst in IR galaxies (like
M 82, Arp 220, NGC 5514, NGC 3256, NGC 2623; see for references
Heckman et al. 1987, 1990; Lipari et al. 2004a,b,c).

The IRAS colour-colour diagrams have been used as an important 
tool to detect and discriminate different types of activity in the 
nuclear/circumnuclear regions of galaxies. Thus, this tool is
also important for the study of possible links between different
phase of galaxy and QSO evolution. Specifically,
L\'{i}pari (1994) found that the IR colours (i.e., IR energy
distribution) of  $\sim$10 extreme  IR +
Fe {\sc ii} QSOs are distributed between the power law (PL) and the
black-body (BB) regions: i.e., the {\it transition area}.
It is important to remark
that of a total of $\sim$10 IR transition objects of this original
sample, the first 4 systems are BAL IR QSOs. Therefore, we
already suggested that BALs IR QSOs (like Mrk 231, IRAS~07598+6508,
IRAS 17002+5153 and IRAS 14026+4341) could be associated with the {\it
young phase of the QSO activity/evolution (and the link between
IR mergers and standard QSOs)}.

Very recently, using our data base of more than 50 IR Mergers and QSOs
with galactic winds and using for comparison the large sample of standard
PG QSO (from Boroson \& Green 1992) we have expanded our previous study.
L\'{i}pari et al. (2005; in their Fig. 15) show the IR
energy distribution [spectral indexes:
$\alpha$${(60,25)}$ vs. $\alpha$${(100,60)}$]
for IR mergers and IR QSO with GW (originally 51 IR systems).
An inspection of this diagram clearly shows the following:
(i) All the IR mergers with low velocity OF (LVOF) are located very close to
the BB and starburst area.
(ii) Almost all the IR QSOs with extreme velocity OF (EVOF) are located in the
transition region.
(iii) The standard QSOs and radio QSOs are located around the PL region.
(iv) All the BAL IR QSOs are located in the transition region, in almost a clear
sequence: from
Mrk 231 (close to the BB area) $\to$  IRAS 07598+6508 $\to$
IRAS04505--2958 $\to$ IRAS 21219-1757
$\to$ IRAS/PG 17072+5153 and IRAS 14026+4341 (close to the PL area) $\to$
standard QSOs.

These results first confirm our previous finding (obtained from
a small sample of IR galaxies): in the sense that
{\it IR QSOs are probably young, composite and transition objects}
(between IR mergers and standard QSOs).
Furthermore, in this IR colours diagram a main evolutionary parameter
is the values of the Out Flow: from IR mergers with Low Velocity OFs to IR
OSOs with Extreme Velocity OFs.


\section{ Explosive Models for Formation and Evolution  of Galaxies/QSOs }

The presence of {\it extreme explosions, OF and galactic-winds}
--associated mostly to extreme/massive star
formation processes-- is  an important component for different
theoretical models  of galaxy and QSO formation and evolution.
More specifically, 3 main explosive models were already proposed:

\begin{enumerate}

\item
Ikeuchi (1981) suggested that QSOs were formed and they exploded mainly at
the cosmologicla redshift Z $>$ 4. The shock waves propagate through
the gaseous medium generated cooled shells (at the shock
fronts). Which are split into galaxies of mass of 10$^{10-11}$ M$_{\odot}$.

\item
Ostriker \& Cowie (1981) have proposed a galaxy formation picture in which
(after redshift 100) small seed perturbation are supposed to collapse, giving
rise to a explosive release of energy from the deaths of the first generation
of stars (Pop. III). This energy drives a blast wave into the surrounding gas.
Thereby sweeping up a shell of shocked material, which eventually cools.
These cool shells are  split into galaxies

\item
Berman \& Suchkov (1991) proposed a hot/explosive model for galaxy
formation. They suggest that the period of major star formation
of protogalaxy (or even giant galaxies) is preceded by an evolutionary
phase of a strong galactic wind. Which is driven by the initial burst
of star formation that enriches the protogalaxy with metals.
Thus this process revert from contraction to expansion.
Specifically, the result of this process is the ejection of enriched
material from the outer part of the protogalaxy, while the iner part,
after a delay of few Gyr, finally contract and cools down to form the
galactic major stellar component.

\end{enumerate}

From the observational point of view,
the presence of multiple concentric expanding supergiant bubbles/shells
in young composite BAL + IR + Fe II QSOs,
with centre in the nucleus and with highly symmetric circular shape could be
associated mainly with giant  symmetric explosive events (L\'{i}pari et 2003,
al. 2005, 2006a,b, 2007). In addition,
an explosive scenario for the origin of some BAL systems (e.g., in Mrk 231)
could explain the SN shape of some BAL light curve variability
(L\'{i}pari et al. 2005, 2006a,b).
These giant explosive events could be explained in a composite
scenario/model: where mainly the interaction between the starburst and
the AGN could generate giant explosive events.
In particular, Artymowicz, Lin, \& Wampler (1993) and Collin \& Zahn (1999)
already analysed the evolution of the star formation (SF) close to  super
massive black hole (SMBH) and inside of accretion disks. They suggested that
the condition of the SF close to the AGNs could be similar to those of the
early/first SF events, where giant explosive processes are expected,
generated by hypernovae (with very massive progenitors: M $\sim$100--200
M$_{\odot}$; see Heger \& Woosley 2002).
In accretion disk, the star--gas interactions can lead to a special mode of
massive star formation, leading to very powerful SN or hypernova explosions.

In order to understand giant explosive outbursts
(i.e., from hypernova, population III of stars, etc), it is required
more detailed theoretical and observational studies.
Very recently,
the dicovery of the most luminous SN 2006GY (in NGC 1260, Smith
et al. 2006) powered by the death of extremely massive star (like Eta
Carinae) and with Type IIn SN properties, strongly support the existence of
extreme explosive events associated with very massive stars.


\section{ Role/Evidence of Explosive QSOs and
a New Explosive Model for Formation and End of Galaxies}

New Gemini/GMOS 3D spectroscopic data of young, composite
and transition BAL + IR + Fe II QSOs: Mrk 231, IRAS 04505-2958,
IRAS/PG 17072+5153, IRAS 07598+6508, IRAS 14026+4341 and IRAS 21219-1757
(Lipari et al. 2006a,b, 2007) strongly support the reality of these giant
explosive processes.

\subsection{ The BAL + IR + Fe II QSO: Mrk 231 }

Using high resolution HST and La Palma/NOT images
we detected --for Mrk 231--  4 nuclear expanding superbubbles
with radius r $\sim$ 2.9, 1.5, 1.0, and 0.6 kpc, plus a starburst
toroid at r 0.2 kpc (Lipari et al. 2005, 20006a, 1994).
For these bubbles, Gemini/GMOS and La Palma/Integral 3D data
(H$\alpha$ velocity field map and 3D spectra)
show in the 4 more external bubbles,
multiple emission line components with low and high OF velocities, of 
$\langle V_{\rm OF }\rangle$  low Vel. $= [-(100,400) \pm 30]$,
 and high Vel. $[-(800,1000) \pm 30]$\,km\,s$^{-1}$.
We suggest that these giant bubbles are associated with the large scale
nuclear OF component, which is generated --at least in part-- by the extreme
nuclear starburst: with giant-SN/hypernova explosions.

In addition, we also
found for Mark 231 that the BAL I system could be
associated with  bipolar outflow generated by the weak/sub-relativistic
jet; and the BAL III system with  a supergiant explosive events (L\'{i}pari
et al. 2005; Punsly \& L\'{i}pari 2005).
The variability of the short lived BAL--III Na ID system was studied,
covering almost all the period in which this system appeared (between
$\sim$1984--2004).
We found that the BAL-III light curve (LC) is very similar to the shape of
a SN LC. Therefore the origin of this BAL-III system was discussed, mainly
in the frame work of an  extreme explosive event.

\subsection{ The BAL + IR + Fe II QSO: IRAS 04505-2958}

L\'{i}pari et al. (2005) proposed a {\it composite 
hyper--wind scenario} in order to explain the very extended blob/shell
(of 30 kpc) found in the new BAL QSO IRAS04505-2958 (this BAL IR-QSO was
discovered using the IR colour-colour diagram: Fig. 15 in L\'{i}pari et al.
2005).
In particular, we have  performed a detailed study of HST images and Gemini
GMOS-IFU spectroscopic data of IRAS 04505-2958
(see Lipari et al. 2005, 2007; Lipari \& Terlevich 2006;
see also Magain et al. 2005).
In general, we found that
IRAS 04505-2958 and Mrk 231 show very similar OF process and properties. Even
both QSOs have  "relatively narrow" --or nini/associated-- BALs
(Lipari et al. 2005, 2006a).
They suggested  that extreme explosions and extreme starbursts
are associated mainly with the interaction between: the QSO and the nuclear
star formation process.

More specifically, we have studied in detail the out flow (OF) process and
their associated structures, mainly at two large galactic scales:
(1) two blobs at radius r $\sim$ 0.2 and 0.4$''$ ($\sim$1.1 and 2.2 kpc);
and (2) an external super/hypergiant symmetric shell at r $\sim$  2.0$''$
(11 kpc). In addition, the presence of a very extended hypergiant
shells at r $\sim$  15$''$ ($\sim$80 kpc)  was analysed.
From the study of the  Gemini data the following main results were obtained:
(i) In general the GMOS data show  strong  emission lines,
in almost all the observed GMOS field: $\sim$ 20 $\times$ 30 kpc
($\sim$3.5$'' \times$ 5.0$''$). Furthermore, multiple emission line
systems were detected, in the regions of the shells (which are aligned
at PA $\sim$ 310: i.e. suggesting a bipolar outflow).
These shells also show emission lines ratios consistent with 
an extreme outflow process and the associated shocks. 
(ii) For the two more internal blobs (at r $\sim$ 1 and 2 kpc) the GMOS
data show that these structures are symmetric shells in expansion, with
{\it very similar} properties to those detected recently in the supershells
of Mrk 231.
In particular, a strong blue continuum component was observed in the region 
of the galactic wind associated with these 2 shells.
(iii) For the external supergiant shell at 2.0$''$ (11 kpc) 
all the kinematics GMOS maps of the ionized gas ([O II], [Ne III], [O III],
H$\beta$, H$\alpha$) show  continuity in velocities between the QSO
and this external shell. The GMOS data suggest that this shell is forming
a satellite/companion galaxy.
(iv) Using  the optical GMOS and HST data plus the IR  observation of
IRAS 04505-2958 we have confirmed that the more probable source of
ultra-luminous IR energy is the QSO.

Therefore, we found that these new GMOS data
are in good agreement with our extreme OF + explosive scenario: where part
of the ISM of the host galaxy was ejected in the form of multiple shells
(which could generate satellite galaxies).
This extreme OF process detected in IRAS 04505-2958 could be also associated
with 3 main processes or steps in the evolution of QSOs and their host galaxies;
specifically:
(i) to stop the accretion process in SMBHs/QSOs; 
(ii) the formation of satellite/companion galaxies by giant explosions; and
(iii) to define the final mass of the host galaxy, and even if the
explosive nuclear outflow is extremely energetic, this process could
disrupt an important fraction  (or even all) of the host galaxy.
Thus this type of giant QSOs explosions is an interesting process in order
to consider as the base for a new model of satellite galaxy  formation 
and (a first) model of galaxy end.


\begin{referencias}

\reference Artymowicz, P., Lin, D., Wampler, E. 1993, \apj, 409, 592 

\reference Berman, V.G., Suchkov, A. 1991, \apss, 184, 169 

\reference Boroson, T., Green, R. 1992, \apjs, 80, 109 

\reference Collin, S., Zahan, J. 1999, A\&A, 344, 433 

\reference Frye, B., Broadhurst, T., Benitez, N. 2002, \apj, 568, 558 

\reference Heckman, T., Armus, L., Miley, G. 1987, \aj, 93, 276

\reference Heckman, T., Armus, L., Miley, G. 1990, \apjs, 74, 833

\reference Heger, A., Woosley, S. 2002, \apj, 567, 532 

\reference Ikeuchi, S.  1981, \pasj, 33, 211 

\reference Lipari, S.  1994, \apj, 436, 102 

\reference Lipari, S. et al. 1994, \apj, 427, 174 

\reference Lipari, S. et al. 2003, \mnras, 340, 289 

\reference Lipari, S. et al. 2004a, \mnras, 348, 369 

\reference Lipari, S. et al. 2004b, \mnras, 354, L1 

\reference Lipari, S. et al. 2004c, \mnras, 355, 641 

\reference Lipari, S. et al. 2005, \mnras, 360, 416

\reference Lipari, S. et al. 2006a, \mnras, submitted, (astrp-ph 0607054)
   
\reference Lipari, S. et al. 2006b, Bol.AAA, No. 49, in press

\reference Lipari, S. et al. 2007, \mnras, in preparation, 
  
\reference Lipari, S., Terlevich, R. 2006, \mnras, 368, 1001
 
\reference Magain, P. et al. 2005, Nat., 437, 381 

\reference Maiolino, R. et al. 2003, \apj, 596, L155 

\reference Maiolino, R. et al. 2004, A\&A, 420, 889 

\reference Ostriker, B., Cowie, .  1981, \apj, 243, L127

\reference Punsly, B., Lipari, S.  2005, \apj, 623, L101

\reference Sanders, D., Mirabel, F.  1996, ARA\&A, 34, 789

\reference Smith, N. et al.  2006, \apj, submitted, (astrp-ph 0612617)

\end{referencias}

\end{document}